\documentclass[%
 reprint, 
superscriptaddress,
 amsmath,amssymb,
 aps, physrev,
]{revtex4-2}

\usepackage{graphicx}
\usepackage{bm}

\usepackage{times}
\usepackage[usenames,dvipsnames]{xcolor}
\usepackage{hyperref}

\definecolor{cream}{RGB}{222,217,201}

\hypersetup{
     colorlinks=true,
     linkcolor=Orange,
     filecolor=MidnightBlue,
     citecolor = MidnightBlue,
     urlcolor=MidnightBlue,
}

\usepackage{listings}
\definecolor{codegreen}{rgb}{0,0.6,0}
\definecolor{codegray}{rgb}{0.5,0.5,0.5}
\definecolor{codepurple}{rgb}{0.58,0,0.82}
\definecolor{backcolour}{rgb}{0.95,0.95,0.92}

\lstdefinestyle{mystyle}{
    backgroundcolor=\color{backcolour},
    commentstyle=\color{codegreen},
    keywordstyle=\color{magenta},
    numberstyle=\tiny\color{codegray},
    stringstyle=\color{codepurple},
    basicstyle=\ttfamily\footnotesize,
    breakatwhitespace=false,
    breaklines=true,
    captionpos=b,
    keepspaces=true,
    numbers=left,
    numbersep=5pt,
    showspaces=false,
    showstringspaces=false,
    showtabs=false,
    tabsize=2
}
\usepackage{todonotes}

\makeatletter
\def\maketitle{
\@author@finish
\title@column\titleblock@produce
\suppressfloats[t]}
\makeatother

\newcommand{\orcid}[1]{}

\newcommand{\mkremoved}[1]{}

\newcommand{\tw}{t_\textrm{w}}
\newcommand{\figref}[1]{\ref{#1}}

\newcommand{\be}{\begin{equation}}
\newcommand{\ee}{\end{equation}}
\newcommand{\bea}{\begin{eqnarray}}
\newcommand{\eea}{\end{eqnarray}}

\newcounter{SIfigures}

\newcommand{\SIref}[2]{\setcounter{SIfigures}{#1}\addtocounter{SIfigures}{-1}\refstepcounter{SIfigures}\label{SI-#2}\ref{SI-#2}}

\newcommand{\captionSIref}[2]{{\color{orange}S#1}}
\newcommand{\rev}[1]{#1} 

\begin{document}

\preprint{APS/XXX}

\newcommand{\TITLE}{%
Time is length in self-similar logarithmic aging of physically \\ cross-linked \rev{semiflexible} polymer networks
}

\title{\textbf{\TITLE}}

\author{Patrick Ilg\thanks{Corresponding author}\orcid{XX}}
\affiliation{School of Mathematical, Physical, and Computational Sciences, University of Reading, Reading, RG6 6AX, United Kingdom.}
\email{Contact author: p.ilg@reading.ac.uk}
\author{Clarisse Luap\orcid{XX}}
\affiliation{Independent Researcher, 8049 Zurich, Switzerland}
\email{clarisse.luap@alumni.ethz.ch}
\author{Martin Kr\"oger\orcid{XX}}
\affiliation{Magnetism and Interface Physics, Department of Materials, ETH Zurich, 8093 Zurich, Switzerland.}
\affiliation{Computational Polymer Physics, Department of Materials, ETH Zurich, 8093 Zurich, Switzerland.}
\email{mk@mat.ethz.ch}

\date{October 9, 2025}

\begin{abstract}
Physical aging in polymers is a fundamental yet poorly understood phenomenon, as diverse macromolecular systems exhibit remarkably similar slow dynamics. Through molecular dynamics simulations of physically crosslinked networks composed of semiflexible polymers, we identify a previously unexplored class of self-similar aging. The network undergoes ultra-slow coarsening characterized by a logarithmically growing mesh size, $L(t) \sim \ln t$, which governs the spatial organization, cohesive and bending energies, and the aging dynamics
of the system. This single time-dependent length scale defines an internal clock, giving rise to spatio-temporal self-similarity of both structure and dynamics -- offering a perspective on aging in soft and disordered materials.
\end{abstract}

\maketitle

Aging in polymeric systems 
is a compelling but ill-understood phenomenon that
plays a crucial role for material properties and 
practical applications \cite{struik_physical_1977,hutchinson_physical_1995,zhang_investigation_2023}.
Increasing relaxation times due to aging lead to
increasing creep compliance in amorphous polymers \cite{struik_physical_1977} and increasing storage moduli for cellulose suspensions \cite{derakhshandeh_ageing_2013}, protein-based biopolymeric gels \cite{manno_amyloid_2010}, or cytoskeletal networks \cite{lieleg2011slow}.
\rev{Here, we explore physical aging in reversibly crosslinked semiflexible polymer networks that results from spontaneous relaxation processes after system preparation in a non-equilibrium initial state, e.g.\ quenching from a high to a low-temperature phase.}
Aging systems fail to reach equilibrium on
experimental time scales and are therefore non-ergodic,
but different polymeric systems show  universal characteristics with rather similar mechanical properties during aging \cite{struik_physical_1977}. 
Corresponding observations have been made in glassy and amorphous systems 
\cite{cugliandolo_dynamics_2002,ruta_relaxation_2017}, for which common mechanisms in terms of power-law waiting times have been proposed \cite{Lomholt2013,amir_relaxations_2012}.
In particular, the weak-ergodicity breaking hypothesis allows a simplified description of aging dynamics in terms of a scaling function.
While this hypothesis and the scaling assumption have been tested for several glassy systems \cite{kob_aging_1997,kob2000fluctuations,park_domain_2012,folena_weak_2023,cugliandolo_dynamics_2002}, only few studies have reported such an analysis for aging polymers \cite{christiansen_coarsening_2017}.

For a large class of systems, aging can be related to a growing length scale due to coarsening.
As predicted theoretically, spinodal decomposition leads to power-law coarsening 
\cite{bray_theory_1994}.
\rev{In this context, 
self-similar coarsening of two-phase mixtures 
\cite{sun_self-similarity_2018} and network-forming systems 
\cite{tateno_power-law_2021} have recently been studied.}
For non-disordered systems like ferromagnetic domains, power-law growth can be related to so-called simple aging \cite{cugliandolo_dynamics_2002,Henkel_Pleimling_Sanctuary_2007}.
For disordered systems on the other hand, thermally activated dynamics suggests a slow, logarithmically growing length scale $L(t)\sim(\ln t)^{1/\psi}$ with positive exponent $\psi$
\cite{huse_pinning_1985}. 
While previous studies found indications for power-law growth, it has been argued that initial transients may mask a crossover to slower growth at later times \cite{park_domain_2012}.
Logarithmically slow coarsening has so far been reported only for a few systems, such as a frustrated Ising model  \cite{shore_logarithmically_1992}, crumpled sheets \cite{shohat2023logarithmic}, and physically crosslinked networks formed by semiflexible  
polymers \cite{kroger_ultra-slow_2025}.

Here, we consider the latter polymeric system: a generic bead-spring model of $1000$ interacting chains, each consisting of $30$ beads, including cohesive energy $E_\textrm{coh}=1.4$ and bending stiffness $\kappa\in\{20,50\}$ (ensuring the formation of a percolated network),
and carefully study its aging properties 
at fixed temperature $T=1$ and number density $\rho=0.05$ (in reduced units)
via molecular dynamics simulation. 
At startup, all semiflexible chains are placed randomly without overlap.
Ensemble averages are performed over 20 independent realizations of the system. \rev{Details of the model and simulations are available in the End Matter section.}
\rev{Self-similar coarsening in these networks we established already in our previous work \cite{kroger_ultra-slow_2025}, showing that network structures at different times are statistically identical when scaled with the coarsening length $L(t)$. 
Here, we study the resulting aging effects and elucidate the role of $L(t)$ for dynamic properties. 
}

\emph{One-time quantities: logarithmic coarsening}--- 
Thermodynamic quantities like the bending energy ($e_a$) and cohesive energy ($e_\textrm{pair}$) per particle are 
one-time quantities that can be defined using a single particle configuration only. 
While such quantities are time-independent in stationary states,
we observe decreasing values of $e_a$ and $e_\textrm{pair}$ with increasing waiting time $\tw$ since system preparation,
indicating ongoing relaxation processes (see Fig.\ \SIref{1}{fig:energies-tw} in Supplemental Material \cite{citeSI}).
Most importantly, these processes are found to become slower and slower as the system ages, such that 
no stationary state is reached
even for very long simulation times 
(as discussed below).

\begin{figure}
    \centering
    \includegraphics[width=0.8\columnwidth]{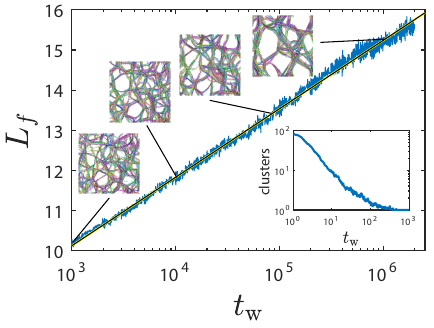}
     \caption{Mean filament length $L_f$ vs. waiting time $\tw\ge t_p$ for $\kappa=50$.
     The black-yellow line shows Eq.\ \eqref{Lf_logtw} with $a=0.74$ and $t_0=0.0012$.
     Inset: Mean number of clusters vs.\ $\tw$ (particles that are bonded via permanent or temporary bonds belong to the same cluster, as in \cite{kroger_ultra-slow_2025}).
         Snapshots show two-dimensional projections at the respective waiting times $\tw$. Each chain has its own color. 
     }
    \label{fig:Lf_logt}
\end{figure}

\rev{For our model system, the relaxation processes reflected in decreasing values of $e_a$ and $e_\textrm{pair}$ correspond to chains becoming straighter and locally more dense.} These processes are related to structural changes within the network which can be monitored 
e.g.\ by the mean filament length $L_f$ which we
calculate from the skeleton network \cite{kroger_ultra-slow_2025} as the mean contour length of edges connecting the skeleton nodes. 
Figure \ref{fig:Lf_logt} shows the extremely slow increase of $L_f$ with increasing waiting time $\tw$ once a single percolated cluster has been established at $t_\textrm{p}\approx 10^3$ (inset).
Its time evolution can be described for both $\kappa$ to a very good approximation by a logarithmic growth law,
\begin{equation} \label{Lf_logtw}
    L_f(\tw) \approx a \ln (\tw/t_0), \qquad \tw\gtrsim t_\textrm{p},
\end{equation}
where  
$t_0$ 
denotes a microscopic reference time.
Snapshots of one sample for selected waiting times $\tw$ are also shown in Fig.\ \ref{fig:Lf_logt}. 
They illustrate the coarsening network.
\mkremoved{The inset of Fig.\ \ref{fig:Lf_logt} shows the evolution of the number of clusters in the system. 
We identify $t_\textrm{p}=10^3$ as the time when all realizations have reached a fully percolated state exhibiting a single, system-spanning cluster.}

Besides the filament length $L_f$, several other lengths can be used to characterize the network, such as the mean filament diameter $d_f$, the mean pore size $r_p$, 
the persistence length $\ell_p$, 
and the mean weighted and un-weighted chord lengths $\ell_1$ and $l_1$, respectively, that measure the distance between two consecutive network-pore interfaces.
Details on the definition of these quantities and their numerical calculation can be found in Ref.\ \cite{kroger_ultra-slow_2025}.
Same as the filament length $L_f$,
the quantities $d_f, r_p, \ell_p$ and $\ell_1, l_1$ also show logarithmically slow growth. 
What is more, the evolution of the network characteristic lengths and pair and bending energies $e_\textrm{pair}, e_a$ all follow that of the filament length.
\rev{Figure \ref{fig-ms-II-lengths-versus-Lf-PRL-revision} shows an approximate power-law
 relation between all these characteristic network lengths.} 
Therefore, the characteristic sizes of the network during the self-similar coarsening are all controlled by the logarithmically growing filament length.

\emph{Two-time correlation functions and aging} -- 
In experiments as well as in simulations, length-scale dependent relaxation is typically studied via the incoherent scattering function.
\rev{To account for the waiting-time dependence of the aging system, we monitor a generalized definition  of its self-part,
$C_q(t,\tw)= N_b^{-1}\sum_{j=1}^{N_b}\langle \exp{(i{\bf q}\cdot[{\bf r}_j(\tw+t)-{\bf r}_j(\tw)])} \rangle$,
where $N_b$ denotes the total number of beads, ${\bf r}_j$ their unwrapped positions and angular brackets represent ensemble averages for fixed $\tw$ \cite{kob_aging_1997,kob2000fluctuations}.}
For isotropic systems, $C_q$ depends only on the magnitude $q$ of the scattering vector ${\bf q}$.
While time-translational invariance of equilibrium dynamics ensures that $C_q$ depends only on the time difference $t$, this is no longer the case for aging systems.
In fact, the waiting time $\tw$ has been identified as the most relevant material parameter in the aging regime of amorphous polymers \cite{struik_physical_1977}.

\begin{figure}
\centering 
\includegraphics[width=0.48\columnwidth]{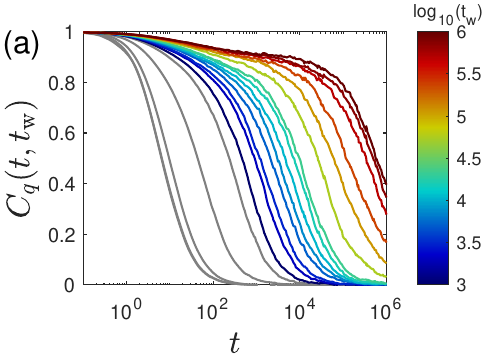}
\includegraphics[width=0.48\columnwidth]{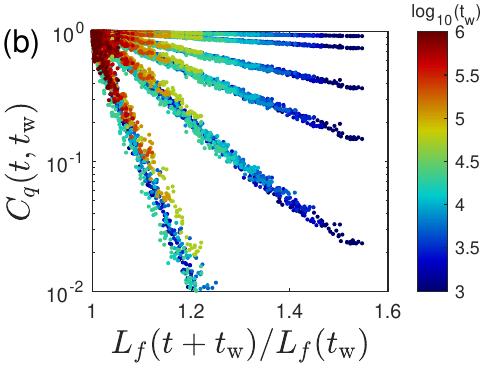}
\\  
\includegraphics[width=0.48\columnwidth]{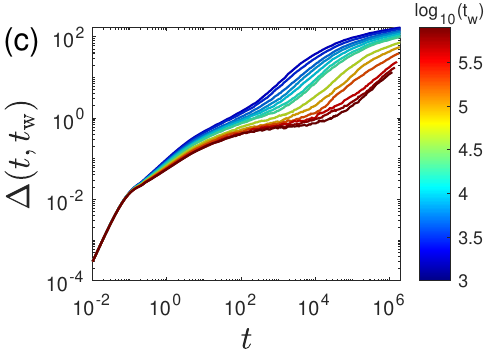}
\includegraphics[width=0.48\columnwidth]{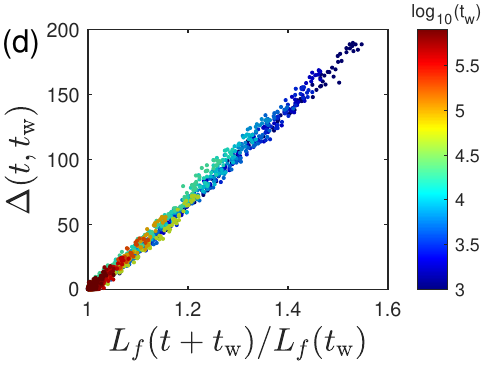}
  \caption{(a) Self-part of the incoherent scattering function, $C_q(t,\tw)$,   
   vs. time $t$ on a logarithmic scale for $q=1$ and $\kappa=50$. 
    Different values of the waiting time $\tw\ge t_p$ are color coded (see colorbar).  
    (b) Same quantity as shown in (a) with the same color code but plotted vs $L_f(t+\tw)/L_f(\tw)$. The wave vector $q$ increases from top to bottom as $q=0.05, 0.1, 0.2, 0.3, 0.5, 1.0$. 
    (c) The MSD $\Delta(t,\tw)$ vs. time $t$ on a double-logarithmic scale (linear plot in Fig.\ \captionSIref{5}{fig:Cq-msd_combined-20b} \cite{citeSI}) for different waiting times (see colorbar). 
    (d) Same quantity as shown in (c) with the same color code but plotted vs. the ratio of the corresponding filament lengths $L_f(t+\tw)/L_f(\tw)$.
  }
  \label{fig:Cq-msd_combined}
\end{figure}

Figure \ref{fig:Cq-msd_combined}-a shows $C_q$ 
as a function of $t$ for different waiting times $\tw$ (color coded).
Relaxations on length scales corresponding to $q=1$ are mainly completed within the time window of our simulations. 
We note, however, that this is not the case for larger length scales corresponding e.g.\ to $q=0.2$ (see Fig.\ \SIref{3}{fig:Cq_t} \cite{citeSI}).
In addition, Fig.\ \ref{fig:Cq-msd_combined}-a shows significantly slower relaxation  with increasing waiting time $\tw$. 
This so-called dynamic slowing down with increasing system age is a typical fingerprint of aging systems.
Figures \ref{fig:Cq-msd_combined}-a and \SIref{3}{SI-fig:Cq_t} \cite{citeSI} suggest a two-step relaxation, where the late-stage relaxation sets in later the larger $\tw$ and the initial relaxation gets suppressed at large length scales. 
Such two-step relaxation
is observed in polymer gels \cite{schupper_multiple_2008} and various amorphous systems \cite{angell_relaxation_2000,ruta_relaxation_2017}.

In the weak-ergodicity breaking scenario for glassy systems, memory of initial conditions is gradually lost such that two-time correlation functions obey a dynamic scaling relation \cite{kob_aging_1997,kob2000fluctuations,cugliandolo_dynamics_2002},
\begin{equation} \label{Cq_ansatz}
   C_q(t,\tw) = C_q^\textrm{short}(t) + C_q^\textrm{age}(h(t+\tw)/h(\tw)). 
\end{equation}
The ansatz \eqref{Cq_ansatz} 
with the short-time behavior $C_q^\textrm{short}(t)$ independent of $\tw$ is in agreement with observations that fast initial 
relaxation in polymers is unaffected by system age \cite{struik_physical_1977}, 
broadly consistent with Fig.\ \ref{fig:Cq-msd_combined}-a. 
The second term, $C_q^\textrm{age}$, 
describes the long-time aging dynamics in terms of a monotonically increasing scaling function $h(t)$ that reparameterizes time.
Different categories of aging systems have been found to share the same scaling function.
This universality in aging is not well understood at present \cite{Henkel_Pleimling_Sanctuary_2007}.
Ferromagnetic domain growth and certain spin glasses show power-law aging,
whereas logarithmic aging is predicted by the droplet model of amorphous systems, with some indications seen e.g.\ in molecular dynamics
simulations of a simple glass former \cite{kob2000fluctuations}.

For aging in spin glasses and liquid-vapor phase separation, the scaling function $h$ was chosen as a dynamical correlation length and mean domain size, respectively \cite{park_domain_2012,roy_aging_2019}.
Since we have already established that coarsening in this system is governed by a single length scale, 
we here use the mean filament length as scaling variable, $h(t)=L_f(t)$.
Figure\ \ref{fig:Cq-msd_combined}-b shows the same data as Fig.\ \ref{fig:Cq-msd_combined}-a together with data for additional $q$ vectors, but plotted vs.\ the ratio of the scaling variable $L_f(\tw+t)/L_f(\tw)$ according to Eq.\ \eqref{Cq_ansatz}.
We observe a very good data collapse for both values of the bending stiffness ($\kappa=20$ shown in Fig.\ \SIref{4}{fig:Cq-msd_combined-20a} \cite{citeSI}) and all wave vectors $q\lesssim0.5$ investigated.
Therefore, the filament length indeed serves as a scaling variable for aging, 
such that
the self-part of the intermediate scattering function can be expressed 
as $C_q^\textrm{age}(t,\tw)\approx \exp{\{a_q[1-L_f(\tw+t)/L_f(\tw)]}\}$, where $a_q$ increases near-quadratically with $q$.
The quality of data collapse worsens for $q\gtrsim 1$, which corresponds roughly to distances smaller than the filament diameter \cite{kroger_ultra-slow_2025}.

\emph{Aging and diffusion}--- 
While diffusion and the mean-square displacement (MSD) are routinely reported for equilibrium and nonequilibrium systems to investigate
their dynamical behavior, aging effects on the diffusive behavior are 
often ignored. 
Some notable exceptions are experiments on aging colloidal glasses
\cite{el_masri_subdiffusion_2010,Boettcher2011,knaebel_aging_2000,lynch_dynamics_2008} 
and tracer particles embedded in
aging polymer networks \cite{sarmiento-gomez_mean-square_2014}.
To capture the waiting-time dependence, we employ a generalized definition of the MSD, 
$\Delta(t,\tw) = N_b^{-1}\sum_{j=1}^{N_b} \langle [{\bf r}_j(t+\tw)-{\bf r}_j(\tw)]^2 \rangle$ \cite{behera_aging_2025,el_masri_subdiffusion_2010}.
We extracted $\Delta$ 
as a function of $t$ for different (color coded) waiting times $\tw$ (Fig.\ \ref{fig:Cq-msd_combined}-c).
All curves are found to coincide in the ballistic regime for short times, $t\lesssim10^{-1}$. 
Consistent with the dynamic slowing down (Fig.\ \ref{fig:Cq-msd_combined}-a), we find that $\Delta$ decreases with increasing $\tw$ for fixed $t\gtrsim 10^0$.
In addition, we observe the build-up of an intermediate plateau with increasing waiting times (Fig.\ \ref{fig:Cq-msd_combined}-c). 
The intermediate plateau in the MSD is typical for many complex and amorphous systems \cite{angell_relaxation_2000,colombo2014self}
and reflects a two-step relaxation mechanism (wriggling/rupture transition) already seen from $C_q$.

Although
the dynamic scaling relation \eqref{Cq_ansatz}  
was originally suggested for other quantities,
$L_f$ can still be used as scaling variable for $\Delta$, as demonstrated by the good data collapse in Fig.\ \ref{fig:Cq-msd_combined}-d for a considerable range of $\tw$ values. 
\rev{In particular, we find an approximate linear relation, $\Delta(t,\tw)\propto  [L_f(\tw+t)/L_f(\tw)-1]$, which can be rationalized from the low-$q$ expansion of $C_q$, i.e., $C_q=1-q^2\Delta(t,\tw)/6+\mathcal{O}(q^4)$}.

\emph{Aging and growing relaxation time}--- 
To better quantify the dynamic slowing down seen in Fig.\ \ref{fig:Cq-msd_combined}, we extract characteristic relaxation times from
fits of the intermediate scattering function (Fig.\ \ref{fig:Cq-msd_combined}-a) to the so-called power-ML function
\begin{equation} \label{powerML}
    C_q(t,\tw) = E_\beta^{\alpha/\beta}(-[t/\tau_q'(\tw)]^\beta).
\end{equation}
The Mittag-Leffler (ML) function with parameter $\beta$ is defined by $E_\beta(z)=\sum_{n=0}^\infty z^n/\Gamma(1+\beta n )$ with $\Gamma(x)$ the Gamma function.
The quantities $\alpha=\alpha(\tw),\beta=\beta(\tw)$ with $0\leq \beta \leq 1$ are treated as fitting parameters in the 
power-ML function \eqref{powerML}, together with the waiting-time dependent characteristic relaxation time
$\tau_q'(\tw)$.
We found the quality of these fits to be very good over the whole parameter range studied, with some deviations only for intermediate $q$ values \cite{kroger_ultra-slow_2025}. 
Equation \eqref{powerML} reduces to 
$\exp{[-(t/\tau_q(\tw))^\beta]}$ for times $t\ll\tau_q'$.
Such stretched-exponential is often used to describe relaxation in complex and amorphous systems, including physical polymer gels \cite{schupper_multiple_2008,cipelletti_universal_2000,rao_bridging_2020}.
The stretched-exponential time $\tau_q$ is related to $\tau_q'$
by $\tau_q'=[\alpha/\beta^2\Gamma(\beta)]^{1/\beta}\tau_q$.

\begin{figure}
\centering
\includegraphics[width=0.48\columnwidth]{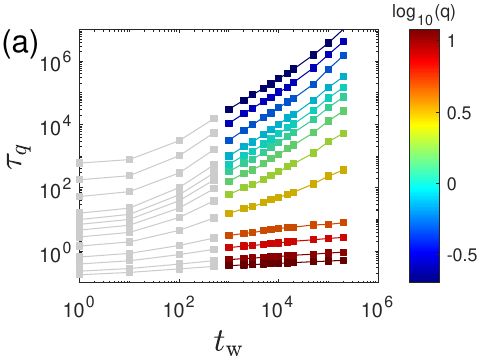}
\includegraphics[width=0.48\columnwidth]{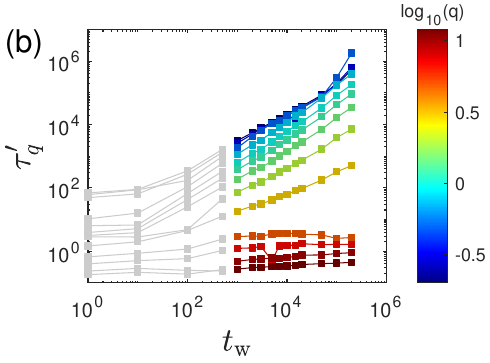}\\
\includegraphics[width=0.48\columnwidth]{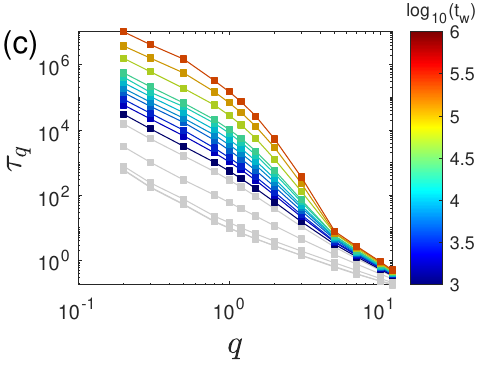}
\includegraphics[width=0.48\columnwidth]{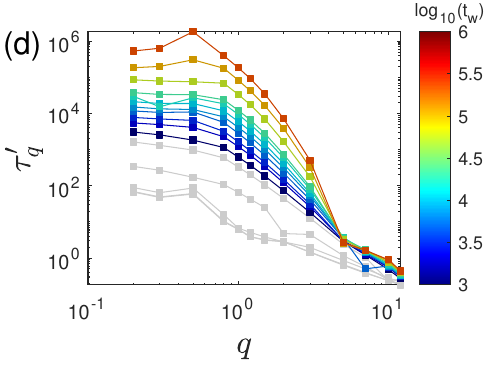}
\caption{(a) The relaxation times $\tau_q$ and $\tau_q'$, respectively, determined from Eq.\ \eqref{powerML} versus $\tw$ for $\kappa=50$  
and different $q\ge 0.2$ (see colorbar). 
(b) The same data are shown versus $q$ 
and different $\tw\le 2\times 10^5$ (see colorbar). 
As before, data for $\tw<t_p$ are shown in dark gray.
}
\label{fig:tauqprime_logq}
\end{figure}

Figure \ref{fig:tauqprime_logq} shows the effect of $\tw$ and $q$ on the characteristic relaxation times $\tau_q$ and $\tau_q'$.
They both increase with increasing $\tw$, as expected for \rev{aging systems}. 
This increase occurs after the network has formed, $\tw\gtrsim t_\textrm{p}$, and is particularly pronounced for small $q\lesssim 2$.
While the relaxation time $\tau_q$ associated with the stretched-exponential  shows a power-law  $\tau_q\sim q^{-\nu}$ for small $q$ values,  
a signature frequently reported for several amorphous and network-forming systems \cite{cipelletti_universal_2000,pastore_elastic_2021,rao_bridging_2020}, 
the relaxation time $\tau_q'$ of the power-ML function approaches a plateau for decreasing values of $q$. 
We find that the plateau is reached for $q$ values 
corresponding to length scales larger than the filament diameter 
, $q<0.05$ (Fig.\ \ref{fig:tauqprime_logq}).
The zero-$q$ plateau values of relaxation times $\tau'_0$ increase strongly with $\tw$. 
We note that these result depend only weakly on the bending stiffness $\kappa$.

The qualitatively different behavior of $\tau_q$ and $\tau'_q$ for small $q$ 
deserves further comments. 
First, $\tau'_q$ is the characteristic relaxation time entering the power-ML function \eqref{powerML} and therefore the primary time scale, at least within this family of fit functions. 
More importantly, $\tau_q$ is determined from $C_q$ 
for times $t\ll\tau'_q$, where the network remains effectively frozen. 
Therefore, values $\tau_q > \tau'_q$ appearing for low $q$ should be interpreted very carefully due to large-scale network relaxation,  
which is 
captured by $\tau'_q$.  
In coarsening systems, it has been argued that  relaxations on scales larger than the domain size are irrelevant or inactive \cite{cugliandolo_dynamics_2002,sun_self-similarity_2018}. 
This argument goes together nicely with the leveling off of $\tau'_q$ for low $q$ seen in Fig.\ \ref{fig:tauqprime_logq}. 

\begin{figure}
    \centering
    \includegraphics[width=0.49\columnwidth]{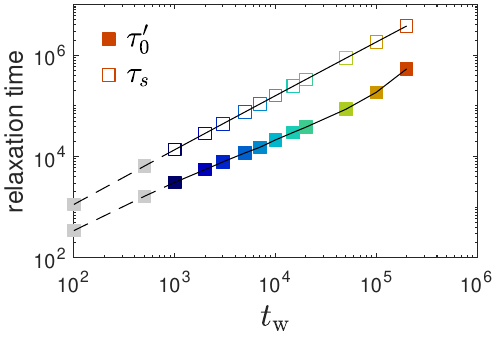}
    \caption{The large-scale relaxation time $\tau^\prime_0$ (full symbols) obtained from $C_q(t,\tw)$ and the separating time $\tau_\textrm{s}$ for coarsening (open symbols) obtained from $L_f(\tw)$ as function of $\tw$ on a double-logarithmic scale.
    }
    \label{fig:taus_logtw}
\end{figure}

To further investigate the links between aging and coarsening dynamics 
we consider the separating time scale for coarsening, $\tau_\textrm{s}=L_f/\dot{L}_f$, where $\dot{L}_f$ denotes the time derivative of $L_f$ 
\cite{cugliandolo_dynamics_2002}.
General arguments suggest that domains are basically frozen-in for times $\tw\ll\tau_\textrm{s}(\tw)$ and substantial coarsening proceeds only for times larger than $\tau_\textrm{s}$.
According to Eq.\ \eqref{Lf_logtw}, the separating time scale increases with system age as 
$\tau_\textrm{s}\approx \tw\ln(\tw/t_0)$.
We find (Fig.\ \figref{fig:taus_logtw}) that $\tau_\textrm{s}$
and the large-scale structural relaxation time $\tau_0'$
show a very similar increase with waiting time $\tw$, emphasizing the strong link between aging and coarsening.
This observation holds for both values of bending stiffness investigated, see Fig.\ \SIref{6}{fig-tau-break}-d \cite{citeSI}.

\emph{Theoretical underpinnings}---
The droplet theory of glasses \cite{huse_pinning_1985,fisher_nonequilibrium_1988}, though developed for disordered systems, may provide a useful framework for understanding some of our main findings and in particular superuniversality of aging in semiflexible networks without quenched disorder.
If coarsening proceeds mainly via filament breakage, typical activation energies scale as 
\rev{$E_\textrm{break} \sim d_f^{\psi/\nu} \sim L_f^\psi$}, 
where the second proportionality follows from self-similarity, see \rev{Fig.\ \ref{fig-ms-II-lengths-versus-Lf-PRL-revision}.}
Crucially, the droplet theory assumes thermally activated relaxation, leading to filament breaking time scaling as $\tau_\textrm{break}\sim \exp(\Upsilon L_f^\psi/T)$ with a constant $\Upsilon$. 
As breakage governs structural relaxation, $\tau_\textrm{break} \approx \tw$, 
the characteristic length grows as $L_f \sim (\ln \tw)^{1/\psi}$. 
We independently verify a near-linear growth of relaxation times with age 
from a network analysis and the change in the number of filaments, which also predicts $\tau_\textrm{break}$ to be proportional to $\tau_\textrm{s}$ (Fig.\ \SIref{6}{fig-tau-break} \cite{citeSI}).
Figure \ref{fig:Lf_logt} suggests $\psi \approx 1$. 
Bouchaud further proposed to define a `glass length' $L_g$ in analogy to the glass transition temperature through the relation 
$\Upsilon L_g^\psi = {\cal A}T$ \cite{bouchaud2000}. 
For our system we obtain $L_g\approx 26$ in Sec.\ \SIref{4}{SI:relaxtimes} \cite{citeSI}, 
meaning networks with filament lengths larger than $L_g$ cannot be equilibrated within a reasonable time frame.  
This length is within a factor of two  of our final configurations
\rev{(still much smaller than our system size)}.

\emph{Concluding remarks}--- Despite lacking quenched disorder, we here demonstrate that \rev{semiflexible polymer} networks largely obey the predictions of the droplet model of glasses, including logarithmic domain growth and superuniversality of aging \cite{huse_pinning_1985,fisher_nonequilibrium_1988}.
While determining domains in amorphous systems is often difficult, the networks under study here are 
defined by a single length scale $L$, which we choose as the filament length $L_f$ \cite{kroger_ultra-slow_2025}.
Since filaments become longer and thicker with increasing waiting time $\tw$, breakage events become more and more rare as the system ages, slowing down coarsening.
The resulting logarithmic, self-similar coarsening goes together with self-similar aging dynamics. 
Consistent with the notion of superuniversality predicted by the droplet model of glasses,  %
we find the growing $L_f$ to serve as a scaling function for the aging dynamics not only for the MSD, but also for the intermediate scattering function for any $q$-value below the filament thickness. 
\rev{Within the network-forming regime, these findings do not depend on the specific value of the bending stiffness.}
\rev{Note that the filament length is also the crucial ingredient in theories of semiflexible polymer networks  \cite{broedersz_modeling_2014,kroy_dynamic_1997}. It remains to be seen whether mechanical properties are governed by $L_f$ as well.}

Contrary to the essentially athermal, stress-induced relaxation advocated for other polymer gels that show anomalous aging \cite{pastore_elastic_2021,cipelletti_universal_2000}, 
our results better align with arguments of 
linear growth of free energy barriers with the size of clusters that have been identified as a key requirement for logarithmic coarsening \cite{shore_logarithmically_1992,shohat2023logarithmic}.
One can argue that the logarithmic growth law is rather robust since the dynamics is hierarchical \cite{cugliandolo_dynamics_2002,sun_self-similarity_2018}, i.e.\ driven by relaxation on length scales up to $L_f$, whereas length scales larger than $L_f$ are mostly inactive. 
This argument is supported by the characteristic relaxation times $\tau'_q$ approaching a plateau value for low $q$ (Fig.\ \ref{fig:tauqprime_logq}) and showing a similar waiting-time dependence as the coarsening time $\tau_\textrm{s}$.
Borrowing the notion of `time is length' from a study of aging in the Edwards-Anderson 
spin glass model \cite{berthier_geometrical_2002}, we find that the internal time and effective age of the network systems under study can be interpreted in terms of the filament length. 
That $L_f$ encodes the effective age of the system provides a convenient way of studying aging dynamics in \rev{semiflexible polymer networks} and glassy systems, but could also open a range of practical applications, e.g.\ with regards to memory and storage. 
\rev{Our study could be relevant for self-assembling biological systems forming fibrous networks such as those involved in the cytoskeleton \cite{broedersz_modeling_2014}, which 
can show significant aging effects  \cite{lieleg2011slow,Schepers2021}.} 
\rev{Our model might also contribute to ongoing research on pathological protein aggregation into fibrous networks, linked to some neurodegenerative diseases and cancer \cite{elbaum2019matter}.
}

\emph{Acknowledgments}---%
P.I.\ was supported by EPSRC via grant EP/X014738/1 and gratefully acknowledges hospitality of ETH Zürich where part of this research was undertaken.

\emph{Data availability}---%
The data that support the findings of this article are openly available from \url{https://github.com/mkmat/FENE-CB-time-is-length}.

%

\section*{End Matter}

\begin{figure}[h]
\centering 
\includegraphics[width=0.49\columnwidth]{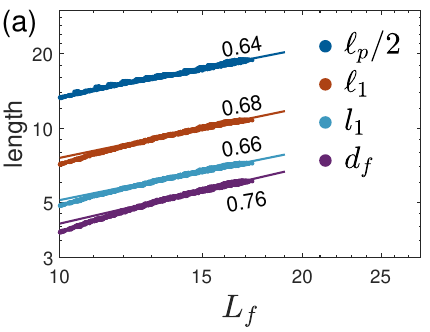}
\includegraphics[width=0.49\columnwidth]{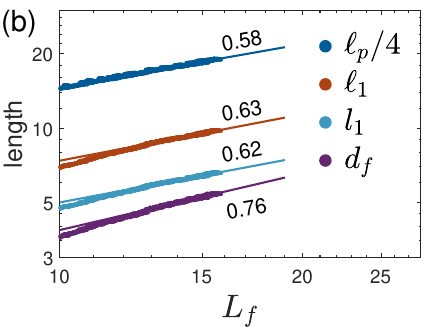}
\caption{\rev{{\bf Characteristic lengths} $L\in\{\ell_p,\ell_1,l_1,d_f\}$ vs.\ $L_f$ (in the course of $\tw$) in double-logarithmic representation for (a) $\kappa=20$ and (b) $\kappa=50$. The exponents $\nu$ in $L\propto L_f^\nu$ have been added to the power-law fits  shown as straight lines.}
}
\label{fig-ms-II-lengths-versus-Lf-PRL-revision}
\end{figure}

\rev{%

\subsection{Model details and cohesive energy} \label{sec:MD}

The model and its implementation using {\small \sf LAMMPS} \cite{LAMMPS} had been described in detail in Ref.\ \cite{kroger_ultra-slow_2025}. 
All multibead, semiflexible chains are initially placed without overlap in a cubic box with periodic boundary conditions at bead number density $\rho=0.05$. Lennard-Jones (LJ) units are used throughout. Permanent connectivity along a chain is ensured through the finitely extensible nonlinear elastic (FENE) potential \cite{Warner1972}, 
$U_{\rm FENE}(b) = -\tfrac{k}{2} R_0^2 \ln[1-(b/R_0)^2]$, 
with parameters $k=30$ and $R_0=1.5$, where $b$ denotes the bond length. In addition, all beads interact through a truncated LJ potential,  
$U_{\rm LJ}(r) = 4 \epsilon_b (r^{-12}-r^{-6}-r_c^{-12}+r_c^{-6})$ for $r\le r_c$. For permanently bonded neighbors, the parameters are chosen as $\epsilon_b=1$ and $r_c=2^{1/6}$. For nonbonded pairs, we here employ $\epsilon_b=3$ and  $r_c=1.359$. Whenever two nonbonded beads approach within $r_c$, they can be regarded as forming a temporary, reversible bond. The corresponding cohesion energy is defined by  
$E_{\rm coh} = (2-r_c^6)^2\epsilon_b/r_c^{12}=1.4$ \cite{kroger_ultra-slow_2025}. 
Chain stiffness is controlled through a bending potential acting on consecutive triplets of bonded beads \cite{FENE-B-1997},  
$U_{\rm bend}(\theta) = \kappa \cos\theta$, 
where $\kappa$ is the bending modulus and $\theta$ the bond angle.  
A schematic representation of the model is provided in Fig.\ \ref{schematic-model.fig}-a, illustrating the FENE bonds, LJ interactions, and the definition of cohesive energy. Both permanent (FENE) and temporary (LJ-based) bonds are indicated. Figure \ref{schematic-model.fig}-b further highlights geometric measures used for characterizing the emerging filamentous networks, such as $L_f$, $d_f$, junctions, and pore sizes, which we extract from the network's skeleton and surface \cite{kroger_ultra-slow_2025}.
The solvent is treated implicitly. Its quality is effectively encoded in the cohesive energy $E_\textrm{coh}$, while dynamics are modeled in the free-draining approximation by including frictional forces on each bead. A Langevin thermostat maintains a constant temperature $T=1$.  

\begin{figure}[h!]
\includegraphics[width=\columnwidth]{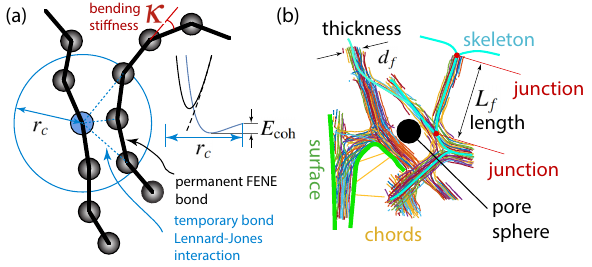}
\caption{\rev{(a) Illustration of the potentials governing the bead--spring chains. Cohesion energy $E_{\rm coh}$ is directly related to the cutoff distance $r_c$, which determines temporary reversible bonds in addition to the permanent FENE bonds. Inset: FENE (dashed black), LJ (blue), and combined potentials (black). (b) Geometric network descriptors based on the polymer skeleton (thinning algorithm), including strand length, thickness, junctions, chord lengths, and pore sizes. Reprinted with permission from \cite{kroger_ultra-slow_2025}.}}
\label{schematic-model.fig}
\end{figure}

\subsection{Network characteristic lengths} 

Our systems show spatio-temporal self-similarity in the sense that not only structural quantities at different times are related by the coarsening length $L(t)$ as in self-similar coarsening \cite{sun_self-similarity_2018,tateno_power-law_2021,kroger_ultra-slow_2025}.  
Also dynamic quantities like $C_q(t,\tw)$ uniquely depend on the ratio of lengths $L(t+\tw)/L(\tw)$, irrespective of the age of the system, $\tw$. 
See Figs.\ \figref{fig:Cq-msd_combined}-b,d where we choose the filament length as the coarsening length, $L=L_f$. These dynamical quantities thus depend uniquely also on $\ln[L(t+\tw)/L(\tw)]$. 
Besides the filament length $L_f$, the percolated networks can be characterized by several other characteristic lengths $L\in\{\ell_1,d_f,\ell_p,l_1\}$. In Fig.\ \figref{fig-ms-II-lengths-versus-Lf-PRL-revision}, we show that these length scales  behave as $L(\tw) \propto L_f^\nu(\tw)$ with a near-constant, $L$-dependent exponent $\nu$, which is (apart from $\ell_p$) sensitive to the definition of the material's surface.  
Since $\ln[L(t+\tw)/L(\tw)] = 
\nu \ln[L_f(t+\tw)/L_f(\tw)]$, these quantities can alternatively be expressed as functions of the ratio of $L$ instead of $L_f$.
Therefore, any of these lengths can in principle be used to characterize the aging dynamics of the network. 
We choose $L_f$ due to its robust definition, good statistics, and its common use in studies of network properties.}

\subsection{\rev{Power-ML fits}}

\begin{figure}[h]
\centering
\includegraphics[width=0.49\columnwidth]{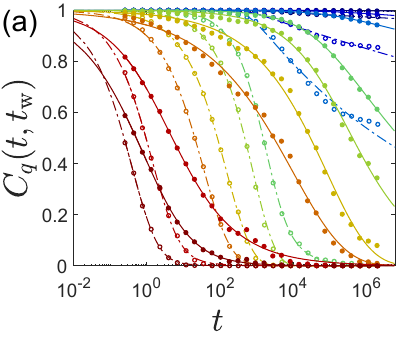}
\includegraphics[width=0.49\columnwidth]{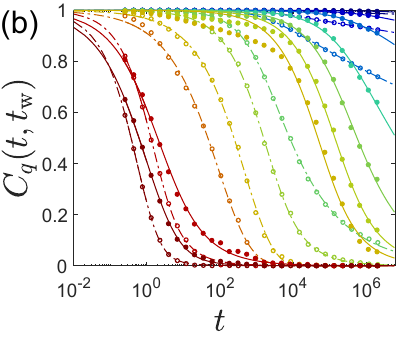}
\caption{\rev{{\bf Incoherent scattering}. The time decay of the self part of the intermediate scattering function, $C_q(t,\tw)$ for selected $q$-values as indicated by color, from $q=10^{-2}$ (blue) to $q=1$ (red)
for (a) $\kappa=20$ and (b) $\kappa=50$.
Open and closed circles correspond to data for $\tw=0$ and $\tw=10^5$, respectively. All lines are
fits to the power-ML function \eqref{powerML}. Panel (b) adapted with permission from \cite{kroger_ultra-slow_2025}.} 
}
\label{fig:Fsqt}
\end{figure}

\rev{Shown in Fig.\ \figref{fig:Fsqt} are fits of the intermediate scattering function $C_q(t,\tw)$ to the power-ML function \eqref{powerML}. The marked difference between short (open circles) and long (full circles) waiting times illustrate strong aging effects with dynamic slowing down. 
Results for different $q$ values can be distinguished by the color code. We observe that fits overall represent the simulation data very well over the whole time window spanning six orders of magnitude, with the exception of intermediate $q$ values for long waiting times where fits overestimate $C_q$ at intermediate $t$. }

\end{document}